\documentclass[journal,12pt]{IEEEtran}
\ifCLASSOPTIONcompsoc
  \usepackage[nocompress]{cite}
\else
  \usepackage{cite}
\fi

\usepackage{hyperref}
\usepackage{url}
\usepackage{graphicx}
\usepackage{booktabs} 
\usepackage{adjustbox}
\usepackage{amsmath}
\usepackage{colortbl} 
\usepackage{xcolor} 
\usepackage[skins]{tcolorbox}
\usepackage[ruled,vlined,noresetcount]{algorithm2e}

\begin{document}

\title{Sparse Architectures for Text-Independent Speaker Verification Using Deep Neural Networks}

\author{\IEEEauthorblockN{Sara Sedighi}\\
\IEEEauthorblockA{Department of Electrical and Computer Engineering\\
Boise State University, Idaho, USA\\
}
\and
\IEEEauthorblockN{Shayan Ramhormozi}\\
\IEEEauthorblockA{Department of Network and Communication,\\
FANAP Telecom\\
}

}

\maketitle

\begin{abstract}

Network pruning is of great importance due to the elimination of the unimportant weights or features activated due to the network over-parametrization. Advantages of sparsity enforcement include preventing the overfitting and speedup. Considering the large number of parameters in deep architectures, network compression becomes of critical importance due to the required huge amount of computational power. In this work, we impose structured sparsity for speaker verification which is the validation of the query speaker compared to the speaker gallery. We will show that the mere sparsity enforcement can improve the verification results due to the possible initial overfitting in the network.

\end{abstract}

\IEEEpeerreviewmaketitle

\section{Introduction}

Recent advancements in deep learning suggested new approaches to train deep networks led to almost human performance level in the image and object recognition, speech recognition~\cite{schmidhuber2015deep,lecun2015deep,maturana2015voxnet,variani2014deep} and data mining~\cite{piergallini2016word,shirvani2016howard}. Approaches based on Information Theory have also been proposed to provide a framework for interpret the deep architecture in a better sense~\cite{shannon2001mathematical}. Some of these new approaches such as dropout~\cite{srivastava2014dropout} will handle overfitting issue~\cite{ioffe2015batch}. For training deep neural networks, network over-parametrizing makes the architecture unnecessarily complicated. Huge computational power is also required for training and model evaluation~\cite{courbariaux2015binaryconnect}.

Up to now, different approaches have been proposed for compressing models. Model compression~\cite{ba2014deep,hinton2015distilling}, pruning~\cite{reed1993pruning,han2015learning}, and $\ell_{1}-regularization$ \cite{collins2014memory} have been proposed so far for this aim. In some previous works such as \cite{denil2013predicting}, it's been declared that training a few portions of the weights is enough by kernel-based estimators. A large amount of the previously performed methods are based on multiple steps of tuning which makes the model hardly scalable. One issue is the model complexity and computational burdon which is related ot the large number of network parameters. Feature selection is one of the approaches for reducing the number of unimportant neurons. Selecting the important features by emoving unimportant elements may impose the weight pruning. A large number of feature methods in this field such as PCA and AEss have been proposed. 

For effective network compression, different methods such as utilizing the group lasso~\cite{yuan2006model}, structure scale constraining~\cite{liu2015sparse}, and Structured Sparsity Learning~(SSL)~\cite{wen2016learning} have been proposed. For most of the research works, there is no evidence of addressing how the accuracy is related to the compression. In this work, we propose the use of sparsity for imposing structured sparsity for speaker verification. We will show that the simple sparsifying the network, can improve the results for speaker verification.

\section{Imposing sparsity}
\subsection{Group sparse regularization}\label{sec:Attention mechanism for group sparse regularization}

We focus on enforcing group sparsity to prune convolutional and fully-connected layers.~Group lasso has widely been used for feature selection by enforcing the sparsity on the weights group~\cite{yuan2006model,meier2008group}. The objective of the group sparsity is to select the effective channels or neurons in case we have a convolutional layer or fully-connected layer, respectively.

Assume a convolutional layer is demonstrated by $W\in R^{C,[Width,Heigth],F}$ and $C$ parameter indicates the input channels, $[Width,Heigth]$ shows the kernel dimension, and $F$ is the number of output channels. The objective loss function is as follows:

\begin{equation}\label{eq:loss-grouped}
\begin{split}
L(W) & = L_{data}(W) + \lambda_{r}.\ell_{2}(W) + \\ & \frac{1}{\sqrt{|G(W^{m})|}}\{\lambda_{gs}.\sum_{m=1}^{N}L_{gs}(G(W^{m})) \}
\end{split}
\end{equation}

\noindent in which, $L_{gs}$ indicates the group sparsity loss. The $|G(W^{l})|$ is the number of channels for the $m_{th}$ layer and $\lambda$ is the hyper-parameter for the loss.~Assume having L different groups of weights, the group sparsity is defined as follows:

\begin{equation}
L_{gs} = \sum_{j=1}^{L}\sqrt{\sum_{i=1}^{|w^{(j)}|}(w_{i}^{(j)})^{2}}
\end{equation}

\noindent in which $w^{(j)}$ demonstrates the $j_{th}$ group of weights and $|w^{(j)}|$ shows the number of weights in the group. 

\begin{figure}[h]
\begin{center}
\includegraphics[scale=0.5]{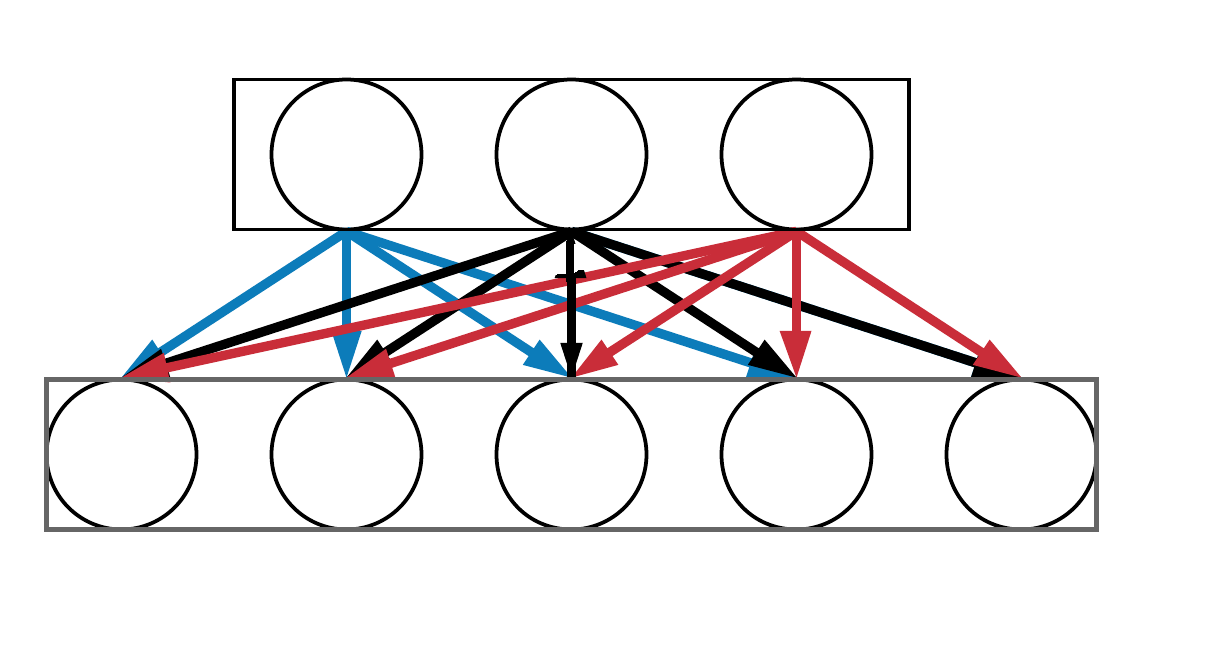}
\end{center}
\caption{Group sparsity on the fully-connected layers.}
\label{fig:loss-grouped-fc}
\end{figure}

The main objective of the group sparse regularization is the removal of redundant features which are activated regarding the network over-parametrizing. For the fully-connected layers, the group is also all the weights connected to the neuron and is shown in Fig.~\ref{fig:loss-grouped-fc}.

\section{Experiments}

The speaker verification is comparing the query speaker to the gallery of speaker models and validate the speaker identity. The speaker verification is mainly divided into text-independent and text-dependent types. In the text-dependent setting, the available spoken utterances are the same. In text-independent setting, however, no assumption is considered for the utterances. The challenge for the latter scenario comes from the fact that the features must distinguishable for both speaker and speech information.

\noindent \textbf{Input:} For each sample sound file, a window of 25ms, with 15ms overlapping is used and the result will be a spectrogram of size $256 \times 100$ for a 1-second duration of the audio sound. For the third dimension, first and second order derivative features are appended together. SpeechPy library has been used for speech feature extraction~\cite{torfi2018speechpy}.\\

\noindent \textbf{Dataset:} We used the VoxCeleb dataset for our experiments~\cite{nagrani2017voxceleb}. There are 1211 available speakers, 40 speakers are chosen for test and the rest are used for training as suggested in~\cite{nagrani2017voxceleb}. The raw audios are extracted from regular Youtube videos which include a variety of internal differences such as background noise and different recording qualities which make the dataset very challenging. For our experiments, we choose very short 1-second utterances with Voice Activity Detection~(VAD) for removing the silence parts. Choosing short utterances have forensics applications and makes our experiments challenging. It is more realistic to consider short utterance because in real-world applications, for most of the times only short utterances of different subjects are available.

For speaker verification architecture, we used convolutional neural networks due to their superiority in applications such as action recognition~\cite{ji20133d}, object recognition~\cite{maturana2015voxnet},speaker verification and audio-visual matching.\\

\begin{figure}[h]
\begin{center}
\includegraphics[scale=0.7]{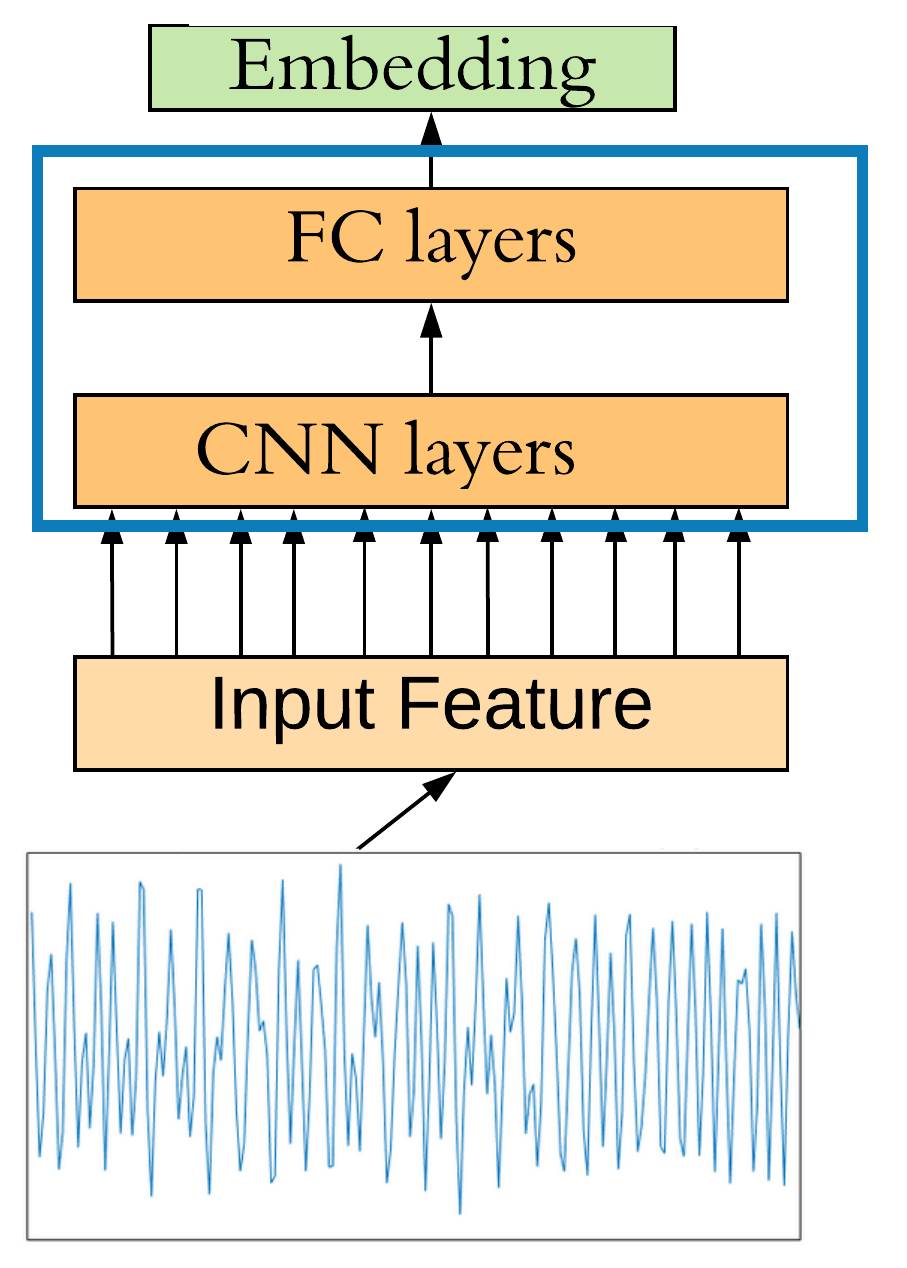}
\end{center}
\caption{The general employed architecture for speaker verification. 
}
\label{fig:architecturespeaker}
\end{figure}

\noindent \textbf{Training and optimization objective:} The architecture is shown in Fig.~\ref{fig:architecturespeaker} which has two deep neural networks with weight sharing. This architecture is a Siamese neural network~\cite{chopra2005learning} and has been utilized for different applications~\cite{sun2018deep,varior2016gated,koch2015siamese}. The main objective of a Siamese network is the creation of a joint output feature space to distinguish between genuine and impostor pairs. The idea is that if two elements of an input pair are coming from the same subject, the output distance should be close by a simple distance metric and should be far away if they have different identities. For this goal, the training loss function should consider both aforementioned conditions. Contrastive loss $L(X,Y)$ is used for that aim and is defined as follows:
\begin{align}\label{eq20}
C(X,Y) = {{1}\over{N}}  \sum_{k=1}^{N} C(Y_k,(X_{1},X_{2})_k),
\end{align}

\noindent where N is the number training samples, and $L(Y_{l},(X_{p_{1}},X_{p_{2}})_{l})$ will be defined as follows:

\begin{equation} \label{eq30}
\begin{split}
C&(Y_{l},(X_{1},X_{2})_{l}) = Y*C_{gen}(D_W(X_{1},X_{2})_{l})\\ &+ (1-Y)*C_{imp}(D_W(X_{1},X_{2})_{l})+\lambda{||W||}_{2}
\end{split}
\end{equation}

\noindent in that $\lambda{||W||}_{2}$ is the regularization parameter. $C_{gen}$ and $C_{imp}$ are the associated costs for the genuine and impostor pairs respectively and will be defined as functions of $D(X_{1},X_{2})$:

\begin{equation}\label{eq40}
  \begin{cases}
    C_{gen}(D(X_{1},X_{2})={{1}\over{2}}{D(X_{1},X_{2})}^2\\
    C_{imp}(D(X_{1},X_{2})={{1}\over{2}}max\{0,{(\eta-D(X_{1},X_{2}))}\}^2
  \end{cases}
\end{equation}

\noindent for which $\eta$ considered to be a predefined margin and $D(X_{1},X_{2})$ is the Euclidean distance between associated output features for the pairs.\\

\noindent \textbf{Architecture:} For the architectures, VGG-Net has been chosen as an effective model for the image classifcation~\cite{simonyan2014very}. The architecture is modified specified to the input features. The output dimensionality of the last layer is set to 64. The average pooling has also been employed for spatial dimension matching~\cite{lin2013network}.

\noindent \textbf{Results and comparison:} The proposed approach will be compared with some other base-line methods as the GMM-UBM method \cite{reynolds2000speaker} which has been selected with 39 MFCCs coefficients including first and second order derivatives. Universal Background Model (UBM) with 512 mixture components has been employed. The I-Vector system\cite{dehak2011front}, is used as one of the methods for the comparison. The results are depicted in Table.~\ref{table:compasison-speaker}. As can be observed, imposing the proposed approach for sparsity outperforms the other methods. 

\begin{table}[h]
\caption[Table caption text]{Comparing the proposed approach with the other methods.}
\label{table:compasison-speaker}
\begin{center}
\addtolength{\tabcolsep}{0pt}
\begin{tabular}{cc}
\toprule 
Model & EER\\
\hline
\midrule
\rowcolor{black!0} GMM-UBM~\cite{reynolds2000speaker} & 28.22 \\ 
\rowcolor{black!0} I-vectors~\cite{dehak2011front}   & 24.91 \\
\rowcolor{black!0} CNN~[baseline]  & 24.68 \\
\rowcolor{black!0} CNN~[SSL]  & 24.11 \\

\bottomrule
\end{tabular}
\end{center}

\end{table}

\begin{figure}[h]
\begin{center}
\includegraphics[width=0.48\textwidth]{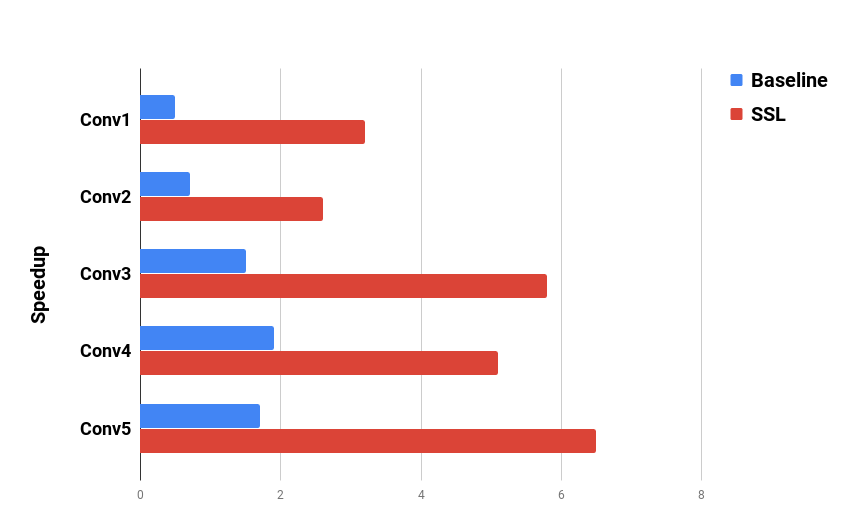}
\end{center}
\caption{The speedup for separate layers.
}
\label{fig:speedup-speaker}
\end{figure}

An important factor caused by enforcing sparsity is the speedup. The results are shown in Fig.~\ref{fig:speedup-speaker}.

\section{Conclusion}

In this work, we proposed the application of sparsity imposition for speaker verification. Experimental results demonstrated the effectiveness of enforcing sparsity due to its potential power for preventing overfitting. This was the direct outcome of removing unimportant elements of the network such as neurons in fully connected layers and output filters in the convolutional layer.

%
%


\bibliographystyle{ieeetr}
\bibliography{ref}
\end{document}